\documentclass[a4paper,12pt]{article}

\usepackage[english]{babel}
\usepackage[T1]{fontenc}
\usepackage{amsfonts}
\usepackage{amssymb}
\usepackage{epsfig}
\usepackage{graphics}

\title{Heterogeneous Strong Computation Migration}
\author{Anolan Milanés, Noemi Rodriguez and Bruno Schulze}

\begin{document}

\maketitle

\begin{abstract}


The continuous increase in performance requirements, for both scientific computation and industry, 
motivates the need of a powerful computing infrastructure. 
The Grid appeared as a solution for inexpensive execution of heavy applications in a parallel 
and distributed manner. 
It allows combining resources independently of their physical location and architecture to 
form a global resource pool available to all grid users. 
However, grid environments are highly unstable and unpredictable. 
Adaptability is a crucial issue in this context, in order to guarantee an 
appropriate quality of service to users. 

Migration is a technique frequently used for achieving adaptation. 
The objective of this report is to survey the problem of strong migration in heterogeneous environments 
like the grids', the related implementation issues and the current solutions. 

\end{abstract}

\newpage
\tableofcontents

\newpage
\section{Introduction} 

\textit{Computation Migration} can be defined as the transfer of a computation from one host to another during 
execution. 
This includes encapsulating and transmiting the computation state (namely, data, 
code and execution state) and restoring it at the destination machine. 
Migration can be done transparently, so the programmer has no control over the migration process, 
or the system may provide some way to control it. 

Discussions about advantages and disadvantages of migration can be found in ~\cite{Eskicioglu90,LO99,CHK94}. 
Although techniques for process migration have been studied 
for several years \cite{MDPW+00,Nuttall94}, it has never been extensively adopted. 
As discussed in~\cite{MDPW+00} and~\cite{CHK94}, this may be due to several
factors, such as performance penalties when compared to alternative solutions,
security issues and sociological factors.
(While sociological factors can be overridden by guaranteeing higher execution priorities or a pre-defined 
degree of resources occupation for the machine owner, they still represent a limitation for the 
adoption of this technique, in particular because security is a real issue.)
The study of process migration was initially strongly motivated by load-balancing concerns in distributed systems.
In this context, migrating a process makes sense only when its remaining execution time is much larger than migration time. However, it is in general very hard to predict the remaining execution time for a running program. This, combined with the relatively high performance costs generally incurred by migration mechanisms, may have contributed strongly to the decrease of interest in the technique.
 
However, the current availability of high speed networks implies in lower penalties 
for the technique. 
Besides, there are other motivations for migration,
in the context of which performance may not be the main issue.
For instance, a computation may be migrated to a node where specific
resources, data, or services are available.
Migration may also be used to fulfill requirements for
fault tolerance and uninterruptible services. 

The emergence of new scenarios, such as those of mobile devices and grid computing,
gives rise to new interest in migration~\cite{LSM00}. 
Grid environments are usually characterized by concurrent execution, domain autonomy, 
resource heterogeneity and high failure probability, which imply in unpredictable resource utilization. 
This motivates the use of techniques that provide adaptation, reliability and maintenance. 
In this scenario, the study of migration seems to acquire new relevance. Furthermore, one important direction of current work in grid computing is that of {\em opportunistic computing}, in which resources are made available for remote users only when local users do not require them. 
Migration is an important mechanism to evict remotely started computations when the machine
owner returns~\cite{TTL05}.
For this reason, we believe it is worthwhile to regain insight into the area by surveying the 
migration techniques that have been proposed.

The problem of migration involves several issues. In this survey we concentrate on the question of {\em how} to move, 
that is, how to implement the transfer of an executing computation. 
This question has very different answers depending on whether we consider that migration will occur
among machines with the same architecture and operating system (homogeneous migration) or among different 
platforms (heterogeneous migration). 
As we have mentioned, grid environments are naturally heterogeneous, and because such environments are 
our motivation, we concentrate, in this study, on heterogeneous migration.


This survey is organized as follows. Section 2 provides background on computation migration, intending to clarify the terms usually employed in the literature, describing various classifications and the problems related to this technique. Section 3 surveys work done in the area of strong heterogeneous migration. 
Finally, we conclude in Section 4 comparing some of the presented systems and discussing the approaches they take for the implementation of this technique.

\section{Preliminaries}

Computation migration consists in moving the execution of a computation from one node to another while preserving its state. The composition of the state depends on the context. For instance, in Unix processes it includes the address space (heap contents, stack, global variables), the execution state (processor state) and the environment information (or resource information, that is, information about open files, but also about messages). 
As another example, the state of a thread running on the Java Virtual Machine (JVM) consists of~\cite{LY99}
the method area (the set of Java classes that includes a Java method currently being executed by the thread),
the object heap (objects accessible from the thread's execution stack),
and the Java stack, organized in blocks called frames. 

In general, migration can be initiated from inside the process (\textit{proactive} or \textit{subjective migration}), 
or from outside (\textit{reactive} or \textit{objective migration}). 
The latter case is usually found in load balancing facilities, allowing the controller engine to command the 
movement of a computation in response to changes on the environment (new resources appeared, 
performance deterioration was detected, etc). 

Migration can be made at process level, usually called process migration, or at thread level (thread migration). 
Migration granularity can be finer: a  single object or a set of objects may move together. 
Emerald~\cite{JLHB88}, for instance, is a distributed language and system designed for the support of object mobility. 
\textit{Mobility} is a term commonly used for referring to migration of objects. 


The term mobility can also be found in literature with other meanings. 
In this report we will use it as a synonym of migration. 
Mobile computation, or computation mobility,
is different from \textit{mobile computing} in that the latter has to do with 
physical mobility (related to physical devices), whereas the first refers to virtual (logic) mobility~\cite{cardelli99}. 
Another term that has been  used in literature referring to migration is Dynamic Software Migration~\cite{ABBC+88} or just Dynamic Migration~\cite{Shub90}. 

It is also important to note the difference between \textit{Computation Migration} and \textit{Code Migration}. 
While code migration involves sending code to some location, computation migration requires the computation
state to be transferred as well.
That is, computation migration requires support for --- but is not equal to --- code migration. 
{\em Mobile agents}, moreover, require support for computation migration and also for transferring authority to 
act on the owner's behalf. 

Mobile agents are code-containing objects that may be transmitted between communicating participants 
in a distributed system~\cite{Knabe95}. 
While process migration is typically initiated from outside the process, mobile 
agents can determine the moment and destination of the migration (although some agent platforms allow 
migration to be also initiated also from outside). 
A mobile agent system is the infrastructure that implements a mobile agent paradigm. 
The agent server, a protected agent execution environment, is responsible for executing agent code 
and provides primitive operations to programmers. 
When an agent requests to be transported to another host, the agent server 
deactivates the agent, saves its state and sends it to the remote agent server, 
which restores and reactivates it.  
Garbage collection is insured by forcing the return of the agent to the creator server after termination. 
Agents are frequently implemented using interpreted languages, because of their features of platform 
independence and dynamic code loading.
Most agent systems have been implemented over the 
Java Virtual Machine (JVM). 

The methods used for implementing migration of processes, threads, or
agents are quite similar.
However, terms used to describe them are often different. 
In this survey, we will employ the term {\em computation migration} to refer indistinctively to 
the movement of any kind of computation (processes or threads) from a source
machine to a target machine, specifying 
the kind only if necessary. 
The moved computation itself will be referred to as computation, executing unit (EUs), as in \cite{FPV98} or 
migration unit, as in \cite{Shub90}. 


Fugetta et al. proposed in~\cite{FPV98} a conceptual framework for understanding code mobility
that has been extensively cited in the literature.
The authors assign a slightly broader meaning to mobility than the one used in the present work:
for them, mobility can be achieved either through migration or through {\em remote cloning}
mechanisms.
Remote cloning basically creates a copy of an Executing Unit at the destination. 
Unlike the case of migration, the original Executing Unit is not destroyed, or detached. 

\begin{figure}[htb!]
\centering
\includegraphics[width=\textwidth]{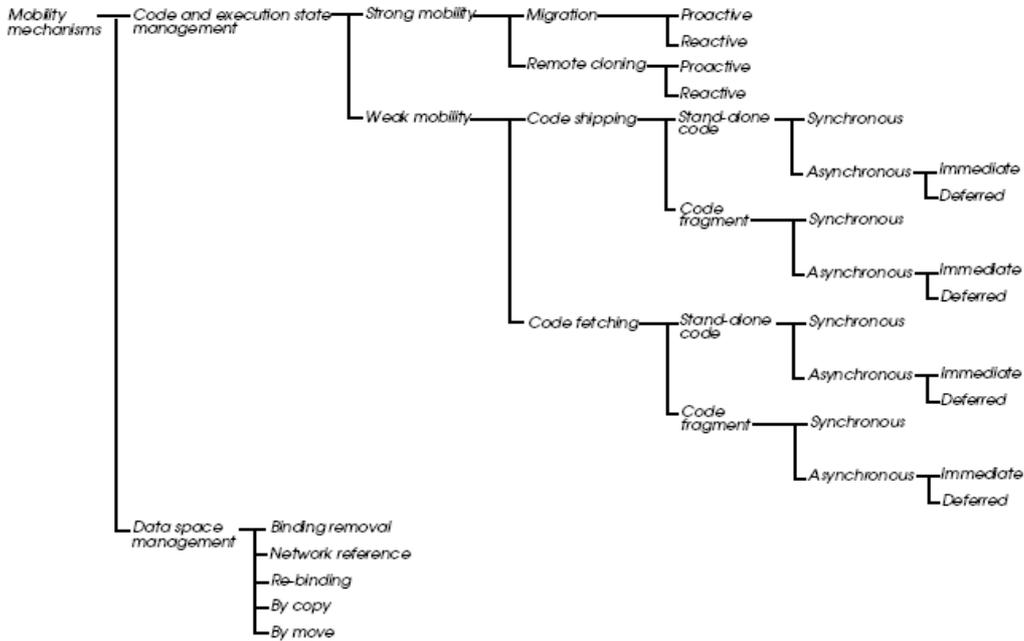}
\caption{Classification of mobility mechanisms \cite{FPV98}}
\label{Fig:classificationMobility}
\end{figure}

Figure \ref{Fig:classificationMobility} shows the classification of mobility mechanisms proposed 
in the work of Fugetta. 
In this classification, mobility can be either {\em weak} or {\em strong}.
Weak migration is the simplest and in consequence, most implemented form of migration. 
In this case, only the code segment is transferred, optionally with some initialization data. 
An example of weak migration in agent systems is provided by Aglets~\cite{LO98}.
After migration, an Aglet will always resume execution from the beginning of the program.
On the other hand, in strong migration, the execution state is also transferred, allowing execution
to restart at exactly the instruction at which it was interrupted at the source host.
NOMADS~\cite{SBBG+00} and D'Agents~\cite{GCKP+02} are examples of agent systems
with strong migration.
Strong migration can be better for programs performing intensive computations and/or long executions, 
but it is harder to implement. 
Because weak migration is extensively implemented,
it makes sense to use it as a basis for providing support for strong migration.

In order to achieve total migration transparency, that is,
for the effects of the movement 
to be hidden from the user and the application, the references to objects and resources 
must also be transferred (open files, etc.). 
Some authors classify the case in which this is handled as 
a third type of mobility called {\em full migration}~\cite{BD01},
but most regard it as a special case of strong migration.
The term \textit{transparent migration} has also been used in literature as a synonym of strong migration, 
whether or not the environment is restored at the target machine~\cite{Funfrocken98,TRVC+00}.
Because of the complexity of this issue, implementations typically 
impose restrictions over the reconstruction of the original environment at the target machine. 
Another problem with transparency is that, while hiding the information about the 
movement of the computation from the programmer certainly reduces programming complexity, it also disallows
the control of errors (maybe caused by network problems or latency) and possible optimizations based on location.

The data space management mechanisms shown in Figure~\ref{Fig:classificationMobility}
address this problem of relocating resources and reconfigurating bindings.
The resources characteristics and the way they are bound to the EU 
restrict the methods that can be used by the data space management mechanisms in each case. 
For example, a huge database can be considered as a non-transferable resource, 
which eliminates copy as a possibility. 
Data space management is orthogonal to the mechanisms that suports 
the mobility of code and execution state. 


To reduce the initial costs of migration,
some of the computation state can be transferred on demand instead of at migration time. 
This technique is called {\em lazy evaluation}. 
{\em Residual dependency} is related to the references the migrated process leaves in the node from where it 
comes or where it was created, typically as a consequence of using lazy evaluation techniques or for 
providing transparency in communications, by redirecting communications through the previously established 
links to the migrated process~\cite{MDPW+00}. 

Migration can be done at {\em user} or {\em kernel} level. 
Kernel level migration modifies the operating system kernel, 
which allows accessing directly the whole state of the process but is complex to implement and 
makes the mechanism dependent of the operating system. 
Implementing migration as a user-level mechanism allows it to be installed with no modification
to the OS kernel.
Besides, typically, user-level implementations are simpler than their kernel-level counterparts, because
of the higher level at which state capture and serialization are done.
On the other hand, operations for manipulating process state are generally not freely available 
at user-level, thus implying in limitations for this approach, and often
higher performance penalties as well.

Given that the motivation of our work is migration in the Grid and similar environments, 
which requires independence of the underlying architecture and the Operating System, 
in this report we  will not focus on migration at kernel level, although techniques for 
process migration in operating systems have been employed in 
higher level migration. 
Methods for implementing kernel-level migration have extensively studied elsewhere~\cite{MDPW+00, Smith88}. 

The concept of handling the state of a process is related to the notion of {\em reflection}
in programming languages, which is the capacity of a program to get and modify information
about its own state at runtime.
Reflective mechanisms allow an executing program to access its computation state and to have knowledge about 
its structure, and also to adapt its behavior as a consequence. 
The implementations of migration can take advantage of reflection techniques.
As an example, Proactive~\cite{BCHV00} uses reflection to choose the method to be executed by the agent at arrival. 
X-Klaim~\cite{BD01} also uses reflection, in this case to capture the agent code to be sent to the remote site.

The problem of migration implies in taking policy decisions to solve the questions of \emph{where} to move, \emph{which} (EU) to move and \emph{when}, and also \emph{how} to implement a mechanism to effectively migrate the process. In this report we examine different techniques for the implementation of mechanisms for strong heterogeneous migration. Homogeneous migration assumes that data representation, machine registers, heap, stack, data segments and machine's data instruction sets are the same. Except otherwise specified, all reported proposals refer to strong migration in heterogeneous systems. 

In fact, when implementing a migration mechanism one must consider
two layers. The inner layer implements state capture, serialization, deserialization, and
restoration. The outer layer manages the problems of mobility at a
linguistic level, which includes the management of locations, etc. Due
to the extent of the problem, this paper studies only the first of these layers.

\section{Heterogeneous Strong Computation Migration}

Migration in homogeneous environments implies in suspending the executing computation,
encapsulating its state, transferring the code and the state information, 
and restarting the computation at the destination using the transmitted information. 
When the source and destination platforms are different, the problem of migration
gets more complicated, because of the need for translation 
of the state of the computation to a format that may be understood at the destination machine.
This is called \emph{heterogeneous migration} and is the typical case in grids,
where no homogeneity assumptions can be usually made. 

Because of the inherent complexity of heterogeneous migration,
a convenient approach is to provide only homogeneous migration 
even in systems running on heterogeneous platforms.
An example is the widely deployed Condor system~\cite{TTL05}.
Migration in Condor is strong but not heterogeneous: the destination node
is chosen among those with the same platform as the source machine.

For the better understanding of the problem of heterogeneous strong computation migration, we 
will subdivide the problem of how to migrate into the problems of the capture/restoration of the state
and that of representing the state to be transmitted.
The capturing and restoring problem involves manipulating information about the internal (local) structure
of the execution state.
The problem of transmissible representations~\cite{Knabe97} is related to determining an
appropriate representation for this information when transferring it between machines.

\subsection {Heterogeneous strong state capture/restore}

A basic mechanism for capturing execution state and restoring it later is to
use the memory image of the computation.
This works up to a point in homogeneous enviroments,
but heterogeneous migration introduces specific requirements,
because of the need of appropriate translations.  
It is no longer possible to transfer memory dumps to the destination with no further modification.
The main issues arise due to differences in instruction sets and data representations, 
which will, for instance, invalidate a Program Counter value from one platform to another.
Even solving a simple representation problem such as endianness 
involves knowing the type and size of the data to be read.  

Data can be translated at the origin, either to a specified architecture or to an  
architecture-independent representation, or alternatively can be translated at the destination.
In this last case, information about data types must be transmitted along with the data itself.

The problem of capturing the structure of data is related to the programming language's
type system.
Most compiled languages keep no runtime information about data types.
Besides, some type mechanisms, such as C's unions, limit the possibility of obtaining type information.
Possible solutions to this issue are to restrict unsafe features of the language or
to modify compilers to deal with these features~\cite{SH98}.
Both can result in non-standard language behaviour.

Much work has been devoted to migrating Java threads~\cite{Funfrocken98,IKKW02,TRVC+00,BHKP+04}.
Java restricts the internal and native information made available
by the virtual machine~\cite{IKKW02}.
The state of Java threads is internal to the JVM: there is no standard API allowing access to it.
Moreover, the state of the stack is also non-portable.
The stack is implemented in most JVMs as a native data structure (a C structure)~\cite{BHKP+04}.
This makes the stack information dependent of the underlying architecture.
A translation step is required to represent the stack state in a platform-independent format
(a Java Object) during the marshalling or serialization process, and the reverse is true for
unmarshalling or deserialization.
This implies in translating the values of local variables and operands to Java values,
which requires access to the type of the values, but Java does not offer this information at runtime.
Runtime type information is embedded in the bytecode of the methods that push the data on the stack.
Techniques commonly used to overcome those problems either have drawbacks on serialization
performance or on portability.

Restoring the computation state consists in creating a new process or thread, reconstructing
the execution state from the transferred state, and restarting execution.
A service must be available to managed the required actions at the destination.
Depending on the implementation platform, the mechanism to restore the stack,
the local variables and the current instruction pointer in every frame can
be more or less complicated.
The execution must resume from the point at which it was suspended at the origin.
But not every language allows jumps to specific points in the code, and even when they do,
translation issues may require the definition of logical points 
marking the next command to be executed.
Java, for instance, facilitates the execution of the received code via dynamic class loading,
but there is no service allowing to restart the execution from the last executed instruction.
The state can be restored by the transferred program itself, by detecting at the
beginning of execution that it must reconstruct its state from a predefined data source,
or by an external service that will restore the whole execution state and
then initiate the program.
Migration based on threads involves the additional issue of synchronization.

Performance penalties due to migration can either be distributed over the execution and
the migration procedure itself, or be concentrated on this last step.
The latter is better for programs where migration is not frequent.

A source code or bytecode instruction may be composed of various machine code instructions, 
thus, when migration is initiated at machine code level, it is necessary to define 
the points at which migration is allowed, to avoid inconsistencies. 
The fact that architecture heterogeneity may cause the program counter location 
to be different at the destination can also be solved by placing logical points, 
acting as labels, to reinitiate the execution at the right instruction, 
given that the migration will only hapen at those points. 
There is a third application for the placement of logical points in the code, 
which is to check for migration requests in the cases in which the system allows 
for objective migration. 
Logical points can be found in literature under different names, like ~\textit{poll points\cite{FCG97}, 
bus stops\cite{JLHB88}, preemption points~\cite{SH98}} or \textit{safe points~\cite{AF02}}, 
and their use was reported in many of the implementations we studied. 
The number and location of those points is a compromise between the performance overhead 
if they are frequently inserted, and the delay in responding to a migration request 
when they are very sparsely distributed. 

The problem of migration is closely related to that of computation
persistency~\cite{Bouchenak01}. Indeed, the persistence of a computation
can be seen as the problem of moving it to the same location, or
otherwise, computation mobility can be achieved
through the restauration of a computation persisted in a different host.
Several techniques for capturing and restoring state are based
on checkpointing facilities~\cite{VD03,AF02}. Checkpointing an
application is the act of saving the application state in persistent
storage
in a form from which it can be restarted later. It is mainly used in
fault tolerance to avoid the need for restarting from the beginning a
process formerly running in a faulty host.

\subsection {Heterogeneous transmissible representations}

After the state is captured, it must be prepared for transmission and 
then transferred to its destination. 
The code and execution state must be transferred in a format, or representation, 
that can be understood by a possibly different architecture. 

The application code may be either transmitted along with the execution state
or obtained, on demand, by the destination host (for instance, by download
from a code base).
The transmissible representation of the code can be either
an architecture-independent representation (to be compiled or interpreted),
or the machine code for the target architecture. 

Recompiling source code at the destination machine guarantees portability
to any platform.
Also, the execution performance of the compiled program will be better than in interpreted schemes. 
On the other hand, there is a delay in restarting due to the recompilation process.

The use of interpreted languages is a valid alternative due to their 
features of dynamic adaptation and portability.
Interpretation allows assuming a homogeneous execution environment, 
supposing that there is an interpreter for every available platform. 
The program to be migrated can be expressed in a platform-independent form, 
as well as the state data, when captured at this level. 
While the implementation of migration using interpreted languages seems to be straightforward 
(migration would consist on the implementation of a mechanism to transmit state data and code), 
common interpreters lack support for execution state capture/restore.

Besides limitations imposed by runtime environments,
there are performance losses inherent to the interpretation procedure.
Some interpreted languages, to improve performance, allow parts
of the program to be coded in a compiled language. 
This is called \textit{dual programming model}~\cite{URI02}, or \textit{interleaving}~\cite{AF02}.
While it can help on the performance point of view, 
it also contributes to the loss of the platform-independence offered by the interpreted approach 
and makes the capture/restoration of the state information more difficult,
as part of it will not be available from the Virtual Machine. 
This kind of application is not usually supported in migration systems. 

If machine code is to be transmitted, it is necessary to generate as many versions of
the compiled programs as the number of platforms that will be supported. 
At migration time, the appropriate program will be selected, if it is not already 
available at the destination. 
This implies in generating a new pre-compiled program for every 
new supported platform, and also in the availability of storage space. 


The Tui system~\cite{SH98} is an
example of the approach based on the migration of native code.
It was built to provide a migration mechanism of Ansi-C programs for four architectures within the 
Unix environment. 
Capture and recovery is carried out with full knowledge of the destination platform, 
and also of the data types and variables used within the program. 
The programs are compiled for each of the four machine types supported by Tui,
producing four different binaries. 
The compiler detects and avoids migration-unsafe features, such as \textit{Unions}, 
to allow the extraction of typing information. 
When the process is selected for migration, a program is called to checkpoint 
the process to an intermediate representation that will be sent to the target machine. 
This program uses the type information generated by the compiler to extract correctly
the data from the executable file.
On the destination, another program takes the transmitted representation and creates a new process. 
After reconstructing all the execution state, the process is restarted from the point 
at which it was checkpointed. 
The system specifies points in the code where the migration is allowable, called preemption points. 

On the other hand, \cite{TH91} describes an implementation of heterogeneous process migration 
based on recompilation. 
In this case, migration involves transmitting a  machine-independent program that, 
when started at the destination, reconstructs the process' state and then continues the normal 
execution of the process. 
This approach has the advantage that it hides the details of data translation in 
the compilers of each machine, but it has the drawback of the increase in the time 
caused by the recompilation and relinking of the program. 
Migration consists of the following steps: suspending the process to be migrated, 
translating the machine-dependent state data to a machine independent representation, 
creating a machine-independent program that represents the process state, 
transferring that program to the target machine, compiling and linking the
transferred program, destroying the source program,
and loading and running the final program on the target machine. 


The Extended Facile~\cite{Knabe95, Knabe97} system, although supporting only weak migration,
takes a hybrid approach to marshalling which is worth mentioning. 
Extended Facile is an extension of Facile, a strongly typed functional programming language 
based on Standard ML with support for concurrency and distribution. 
Extended Facile supports both architecture-independent and machine code 
representations, and the joint transmission of several representations, 
allowing the programmer to choose the representation best suited for its agent. 
It allows the program, for instance, to choose a machine code implementation 
when the destination host has the same architecture as the origin. 

\section{Classification of heterogeneous strong migration techniques}

Methods for the implementation of strong heterogeneous migration are basically similar 
in different environments. 
They consist in a mechanism for capturing the execution state information and saving it 
in stationary or transient storage, and next, transmitting the saved status 
and restarting the saved computation at the remote location. 

Regarding their approach to transmissible representations of code, they can be divided 
between those which use an architecture-independent representation and those which 
use machine code. 
Architecture-independent representations can use either interpretation
or recompilation.

On the other hand, a general classification of the methods used to implement state capture 
and restoration is not quite clear. 
Bouchenak~\cite{BHKP+04} proposed a classification for approaches to capturing the execution
state of Java threads.
We believe this classification can be applied in the more general problem of heterogeneous
state capture.

The classification originally proposed by Bouchenak identifies 
an application-level approach and a JVM-level approach, 
which we generalize to the following:

\begin{enumerate}

\item User program pre-processing: Consists of introducing fragments of code 
(automatically or not) into the user program, in order to make it capable of 
auto-saving/restoring its status. 
This implies in runtime and space overhead caused by the inserted code.

\item Platform modification: Consists of modifying the underlying platform or 
virtual machine to make it provide the required data and functionalities 
necessary to achieve migration.
\end{enumerate}

In Section~\ref{other}, we will mention another approach, 
based on languages which provide some built-in facilities for state capture.

Across the following subsections, we survey systems that were implemented according
to each of these approaches.

\subsection{Code pre-processing, or application-level approach}

In this approach, the user program is modified by
inserting fragments of code that allow the program to save its execution
state and restart the computation by itself.
Such modifications can be done to source or compiled code.
In the case of interpreted languages, the compiled code would
be the pre-compiled code (hereafter called bytecode).

Ferrari et al.~\cite{FCG97}, in one of  the earliest works discussing heterogeneous state capture/restoration,
propose a mechanism called \textit{Process Introspection}.
This consists on pre-processing the application source code (written in Ansi-C) 
to incorporate autonomous checkpoint and restart facilities. 
The target application domain are scientific applications, 
whose high performance requirements would not allow the 
use of interpreted languages for generating platform-independent checkpoints. 
The implementation consists of a library (the Process Introspection Library, PIL), 
and a source-code generator called APrIL (Automatic application of the PRocess Introspection Library), 
which automates the implementation of state capture and recovery. 
PIL provides a mechanism for describing, saving and restoring data values, 
and an event-based mechanism for coordinating the capture/restore activities. 
It can be automatically applied by APrIL to incorporate capture and restore 
functions for platform-independent modules, that is, those written in a high-level language, 
that are type safe, and do not rely on the underlying features of a particular 
platform for correctness~\cite{FCG97}.
The transformation consists in adding prologues to every function, 
which include calls to PIL to register all the local variables or parameter addresses 
found in the function body in the local variable table.
The compiler inserts {\em poll points} followed by code which checks migration requests.
For the cases when automatic transformation is not possible, 
the user can make use of the library directly. 

The states traversed by a migrated process are:

\scriptsize
$Normal Execution \rightarrow State Capture \rightarrow State Recovery \rightarrow Normal Execution$
\normalsize

\begin{figure}[htb!]
\centering
\includegraphics[width=\textwidth]{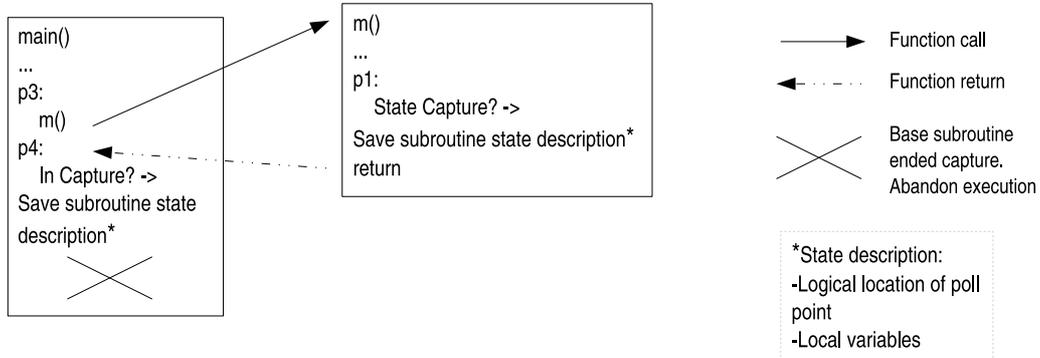}
\caption{Native subroutine return mechanism: State Capture}
\label{Fig:StateCapture}
\end{figure}

The capture of stack state is made through a {\em native subroutine return} mechanism 
(see Figure \ref{Fig:StateCapture}), which consists in saving the state of the 
current procedure (including the logical location of the poll point at which the 
current frame was saved and the local variables and parameters of the function) 
and returning, and is executed recursively until the base routine is reached. 

\begin{figure}[htb!]
\centering
\includegraphics[width=0.7\textwidth]{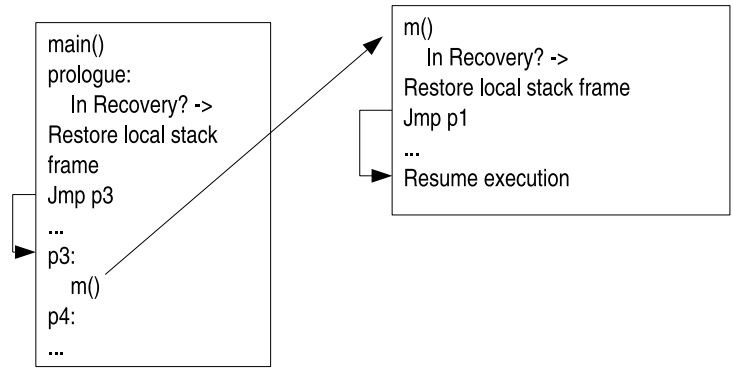}
\caption{Native subroutine return mechanism: State Recovery}
\label{Fig:StateRecovery}
\end{figure}

For restart, every subroutine recovers the data for its local frame and jumps to the 
poll point at which the stack of the current stack frame was captured, as shown in 
Figure \ref{Fig:StateRecovery}. 
For this mechanism to work, 
there must be a poll point in the program after each subroutine call.
Since this mechanism is specified at a platform-independent level, 
the captured state is also valid in any platform, assuming the associated 
data is stored in a universally recognizable format 
for masking issues such as those of data representation. 

Fünfrocken proposes a similar approach~\cite{Funfrocken98} for saving and restoring the state of 
Java programs, using exception handling. 
A compiler inserts try-catch statements in the source code to save the state of execution 
of each thread in different Java backup objects. 
On restart, all the threads that were active at the time of the capture are 
restarted by creating new threads which will be initialized with the contents 
of their respective saved counterparts. 
Synchronization issues emerge here, due to the impredictability of the situation of each
thread at the moment of state capture.
The proposal addresses only subjective migration, 
and thus the points at which migration is to occur are explicit in the code.
The solution was to provide the programmer with a method {\em allowGo} to signal that 
the calling thread is ready to save its state. 
State saving only occurs if all threads have called this method or have initiated state saving. 

The serialization of Java threads in the Brakes
framework~\cite{TRVC+00} is achieved through the instrumentation of
bytecode. 
The technique employed for thread stack capture is very
similar to the \emph{native subroutine return} mechanism proposed by \cite{FCG97}.
In the {\em state capture} state, every method saves the current frame in the Context object 
and returns to the previous method recursively until the end of the stack is reached. 
Given that the instruction counter is not available in Java, a counter was created to represent the 
``last performed instruction'' (LPI). 
On recovery, the thread restores the first frame from the context stack, 
removes it from the context and moves the program counter to the saved LPI 
(using the goto instruction, available only at bytecode level), 
which could correspond to a method being called at the moment of the checkpoint. 
In this case, the method will be called and it will recover its frame from the 
context, in the same way as the previous method did before. 
State verification is done by code inserted automatically by the
bytecode transformer at the beginning of every invoked method.

The MAG grid middleware proposal~\cite{FS05}, an extension of the Integrade middleware for the execution of Bag-of-Tasks applications, is built over the Brakes framework~\cite{TRVC+00} and the JADE framework. 
Its goal was to provide for strong migration, thus portability was a major requirement. 
MAG mobile agents can migrate strongly, objectively and subjectively: 
the Brakes framework was modified to allow the migration to be initiated by an external entity. 
The programmer can also insert new points where the checkpoint may occur. 
The published implementation didn't include the migration of external dependencies 
or multi-threaded applications. 

In JavaGo~\cite{SMY99}, strong mobility is provided by source code transformations.
An exception mechanism is employed for capturing the execution state:
exceptions are thrown recursively until the whole call stack is saved as a chain of State objects. 
Because Java does not include a {\em goto} statement, these transformations
must resort to switch-case statements. This requires some pre-processings (including splitting expressions with side effects) and unfolding techniques. An {\em undock} statement marks the part of the code that will be migrated. 
At restart, unfolding techniques must be also used for loops. 

MobileScope~\cite{MPOY04} is a programming language which supports both weak and strong migration. 
Strong migration is supported by the integration with the JavaGo framework.
MobileScope extended JavaGo to support triggering the migration externally. 
One important aspect of MobileScope is the provision of full mobility, 
by means of channel mobility, 
which allows modifying the resource bindings at runtime. 

Bettini and De Nicola~\cite{BD01} propose a technique for implementing strong 
mobility through a weak mechanism, adaptable
to any language with support for transmitting data and code.
The authors exploit this technique in the implementation
of the X-Klaim programming language. 

In Aglets~\cite{LO98}, the event model allows the programmer
to save the state before migration.
This has been exploited in MobiGrid~\cite{BG04} which focuses on long sequential applications 
and opportunistic computation. 
In MobiGrid, the support for strong migration is achieved by extending Aglets 
to allow the programmer to save the state of execution periodically,
allowing it to be restarted later.  
Drawbacks of this method are the required programmer effort and the fact that the 
execution flow is split into several callback procedures, 
making it hard for the compiler to perform optimizations. 
The fact that the mobility model in Aglets, and, in general, in weak mobile systems, 
forces the programmer to manage the objects to be transferred during migration 
affects the transparency of the process. 

Bytecode instrumentation offers better portability than the 
Virtual Machine modifications techniques and better performance 
that the source interpreted code approach~\cite{BHKP+04}.
A drawback originated by the bytecode-based implementation is the need of 
code maintenance.
There is no guarantee of backward bytecode compatibility in Virtual Machine upgrades, 
so it may be necessary to modify the implementation after JVM upgrades. 
Working over the bytecode (for the case of implementations over Virtual Machines), 
instead of over the source code, has the advantage of the wider instruction set 
and also a better performance. 
Also, for pre-processing, the source code must be available, which is not the case
with libraries and legacy code.
A problem common to both approaches is that the execution state cannot be captured 
in all situations~\cite{BHKP+04}. 
In general, shortcomings of code pre-processing are that it implies 
in changes to program flow and in some 
time and space penalties at runtime, caused by the inserted code. 


\subsection{Modifying/extending the Runtime System}

Another alternative for introducing support for state capture and restoration
is to modify or extend existing platforms.
In this case, there is no need to modify the user code
eliminating the requirements of code availability and lowering 
the execution and space overhead. 
On the other hand, portability is affected.

Agbaria and Friedman~\cite{AF02} propose a transparent mechanism for checkpoint/restart 
in heterogeneous environments focusing on fault tolerance. Their approach is to checkpoint the application state at virtual machine level. The implementation was done over the OCaml virtual machine (OCVM). Since the systems operates at VM level, the checkpoint is only allowed at points where the state of the application is consistent, that is, between instructions or during an instruction which does not modify the state of the system. Whith that in mind, the interpreter checks for a flag signalling a checkpoint request before fetching a new instruction. State capture is based on the tagging of data types and the garbage collection features of OCaml. File-descriptor checkpointing is based on the OCaml support for I/O interception. A multi-threaded consistent checkpoint is achieved by taking the checkpoint only after stopping all threads.

Agbaria and Friedman's work assumes that failures are rare, and that, on that account,
it is preferrable to penalize restart, maintaining checkpointing overhead minimal.
Thus, data is saved in its native representation and translated, if necessary,
when the application is restarted. In order to avoid blocking the application during checkpoint, a new process is forked to save the state and then exits. 

D'Agents~\cite{GCKP+02} (formerly called Agent Tcl) is a mobile agent system with support 
for subjective strong migration. 
Initially implemented in Tcl, it currently allows mobile agents to be written in Tcl, 
Java, and Scheme, and supports strong mobility in Tcl and Java with "significant" 
modifications to the respective interpreters. 
Migration in D'Agents is accomplished with the {\em agent\_jmp} command. 
{\em agent\_jmp} captures the internal state of the Tcl script and transfers it 
to the destination machine, where the execution continues from the next command.
The D'Agents server is multi-threaded: every agent runs in a separate thread.

Bouchenak et al.~\cite{BHKP+04} present a solution for Java thread serialization/deserialization
which intends to build thread mobility or 
persistence while avoiding the performance overhead incurred by previous approaches. 
This is achieved through type inference and dynamic de-optimization techniques. 
The Java Virtual Machine is extended to capture the state of a Java thread as an object 
and to initialize a thread with a particular state. 
The Java compiler was not modified. 
Thread serialization does not handle problems of object sharing 
between threads, distribution, synchronization,
or the management of object dependencies.

The authors compare two prototypes, based on Interpreter-based serialization (ITS) and 
on capture time-based thread serialization (CTS). 
The first was based on the modification of the interpreter to capture the data type 
every time a bytecode instruction saved data on the stack. 
For CTS, the type inference was done by analyzing the bytecode only at thread serialization time. 
The work concluded that given that the serialization of the interpreter (ITS) 
is not compliant with JIT compilation, it would not be a realistic solution. 
It also introduces an execution performance overhead. 
On the other hand, CTS avoids any execution performance overhead, 
but the cost is transferred to the serialization latency and thus it is not advisable 
for applications with a high serialization frequency, such as mobile agents. 
Application-level Java thread serialization would probably be the best solution for those cases.



Illman and others~\cite{IKKW02} propose the use of the Java Platform debugging interface 
to achieve transparent migration in the context of the CIA project 
(Collaboration and Coordination Infrastructure for Personal Agents), 
which deals with the development of an infrastructure for software agents. 
The Java debugging architecture provides access to runtime information like stack frames, 
local variables and the program counter. 
It is possible to stop and resume execution, execute single bytecode instructions 
and set/unset breakpoints. 
A problem in this proposal is how to reestablish the program counter after 
the whole state has been transferred. 
Given that the Java debugging architecture does not include this feature, 
the solutions proposed are the modification of the JVM, or else, 
the instrumentation of the byte code, and take us back to the 
previous approaches. 

\subsection {Migration with Language Support}
\label{other}

Functional languages facilitate the manipulation of functions, allowing their 
transfer to remote hosts for execution. 
Heterogeneous strong migration can be implemented in those systems by means of continuations, 
if those are offered as transmissible structures.
Tarau and Dahl\cite{TD01}, for instance, use this strategy to implement
strong heterogeneous migration in BinProlog. 

Other authors have dealt with the problem of migration using functional languages. 
An implementation for strong mobility over mHaskell~\cite{DTL03} is reported
in \cite{DTL06}.
It is based on weak mobility, higher-order channels, and first-class continuations, 
without the need for changes to the run-time system or built-in 
support for continuations. 
Unlike other proposals, the implementation is based on {\em Monads},
available in Haskell and in other purely functional languages. 

Systems like ARA, NOMADS and Telescript implement support for strong migration by creating new platforms that offer all the required information to execute the procedure. A comparison among those systems is presented in \cite{GCKP+02}. 

\section{Discussion and Conclusions}

Cardelli in Mobile Computation~\cite{cardelli97}, commenting the fact that traditional languages 
and traditional compilers are not well suited for network computing, 
asserts that languages that are not portable on-line will be abandoned 
because they don't provide mobility.
But what support should a run-time engine provide for heterogeneous strong migration?

We have identified two layers for the implementation of migration mechanisms.
The inner layer implements state capture, serialization, and
restoration. 
The outer layer manages the problems of mobility at a
linguistic level.
At the inner layer, we believe the following items should be considered.

\begin{itemize}
\item A mechanism to capture and serialize the state of the execution (including data types, program counter);
\item A mechanism for transferring a computation (data, code and state of execution); 
\item Support for deserializing and restoring the computation (it means, a way to transform back the captured values from the independent representation, and also a way to restart execution from the point where it stopped);
\item The performance offered by these mechanism must satisfy the application goals.
\end{itemize}

Generalizing from the surveyed work, the mechanisms for capture and restore can
be grouped in two major categories, according to the support provided by the programming language.
In the case when there is no language support for state capture/restoration, the 
problem can be solved in two ways, which are based on:

\begin{enumerate}
\item User program pre-processing: Consists in introducing fragments of code (automatically or not) 
into the user program in order to make it capable of auto-saving/restoring its status. 
This implies in runtime and space overhead caused by the injected code.
\item RTS modification (or extension): Consists in modifying or extending the underlying platform 
or virtual machine, seeking to provide the required data and functionalities necessary to achieve migration.
This usually results in better performance but implies in less portability and maintanability.
\end{enumerate}

Language support for state capture/restoration can be expressed either by means of mechanisms such as first-class
continuations, typically offered by functional languages, or by explicit primitives for state manipulation.

The use of continuations has the advantage of enabling strong migration to be implemented
using weak mobility. 
Nevertheless, continuations are almost exclusively supported by functional languages, which have
well known performance limitations.
In general, features commonly present in the so called dynamic programming languages 
like \textit{introspection}, \textit{continuations}, and the possibility of modifying programs 
at runtime make them interesting vehicles for the implementation of migration.
On the other hand, languages and systems that do not offer enough support for state capture 
and restoration force the implementor to choose between preprocessing techniques or modifying the platform.
This  implies in a compromise between portability and performance.

The issue of transferring the computation involves the transmitting the code and captured state
in a way that is understandable at the target node.
The code can be transferred either as
an architecture-independent representation (to be compiled or interpreted),
or as machine code for the target architecture.
The transmission of the captured data seems to be well handled by current solutions
such as Java's RMI.

Table~\ref{TAB:generalization} shows the classification we propose for the case of the surveyed works.

\begin{table}[hbt!]
\begin{center}
\renewcommand{\arraystretch}{.9}
\scriptsize
\begin{tabular}{|l|c|c|} \hline
\bf Name &  \bf Capture/Recovery Method & \bf Code Representation \\\hline
Tui & RTS modification & Native Code \\\hline
Recompilation & RTS modification & Source Code \\\hline
Process Introspection &  Program Transformation & Native Code \\\hline
JavaGo & Program pre-processing & Interpreted \\\hline
Brakes & Program pre-processing & Interpreted \\\hline
Bouchenak & RTS modification & Interpreted \\\hline
CIA project & Debugger Platform & Interpreted \\\hline
D'Agents & RTS modification & Interpreted \\\hline
DTL06 & Language Supported & Interpreted \\\hline
\end{tabular}
\normalsize
\caption{Evaluation according to classification}
\label{TAB:generalization}
\end{center}
\end{table}

Performance is still a issue in the implementation of migration.
However, as we mentioned in the introduction, current motivations for migration,
such as evicting processes in opportunistic computing systems,
make us evaluate migration performance from a new perspective.

Is migration a good idea? It certainly is, for a number of applications and in specific conditions. 
Is strong migration a good idea? Actually, given the costs implied in the mechanisms for adapting it 
to a heterogeneous environment, it seems that a combination between weak and strong migration 
(if possible using weak migration) would be the best answer for systems requiring to use this technology. 
It will depend strongly on the focus of the application: 
if it is a long lasting and high performance application it will require a fault tolerance service, 
thus the need for checkpointing, and high migration performance. 
It is probably not the case for an agent sniffing for information on the internet. 
Further work must be done to evaluate the advantages of the implementation of migration for 
the current potential applications. 
New solutions for security are currently being investigated, like the possibility of 
code signing mobile agents. 
Error management issues should also be better studied.


\section*{Acknowledgements}
This work has been partially supported by CNPq Brasil and PCI/LNCC.

\bibliographystyle{ieeetr} 
\bibliography{survey}

\end{document}